# Using Nagios to monitor the Telescope Manager (TM) of the Square Kilometre Array (SKA)

*Matteo Canzari[1], Matteo Di Carlo [1], Mauro Dolci [1] and Riccardo Smareglia [2]*
*[1] INAF - Osservatorio Astronomico di Teramo, Teramo, ITALY*
*[2] INAF - Osservatorio Astronomico di Trieste, Trieste, ITALY*

## Abstract

SKA (Square Kilometer Array), currently under design, will be a huge radio-astronomical facility, whose management will be performed by a suite of software applications called Telescope Manager (SKA TM) via the TANGO framework. In order to ensure the proper and uninterrupted operation of TM, a local monitoring and control system (TM.LMC) is being developed, with the goal to perform monitoring, lifecycle control and fault management of TM. For the monitoring activity, central in TM.LMC, Nagios (automated by the lifecycle management tool Chef) has been proposed as main toolkit to check resources, services and status of every TM application both at generic and performance level: for this latter purpose, a custom agent has been developed. This led to an integrated fault management module, based on Nagios-Chef integration, which can efficiently handle any abnormal situation

## 1. Introduction

The Square Kilometre Array (SKA) Project is aimed to build a radio telescope that will enable breakthrough science and discoveries, not possible with the current facilities, over the next 50 years. In the overall SKA architecture, two telescopes (SKA MID and SKA LOW) are composed each one by several Elements covering all required functionalities, e.g. DISH and LFAA (the front-end Elements for direct radiation detection), CSP and SDP (data processing and delivery), SAT, SaDT and INFRA (support functionalities). The global orchestration of this huge system is performed by a central element called Telescope Manager (TM) ([3], 2016) which has three core responsibilities: management of astronomical observations (proposal and scheduling), management of telescope hardware and software subsystems (observation execution) and management of telescope engineering data. TM is a complex (and distributed) system, mostly composed by software packages (TELMGT, OBSMGT, LMC sub-elements), web applications and user interfaces running on a hardware e virtualization software platform provided and managed by LINFRA. Each TM sub-element is in turn composed by software (applications) and data (configuration data, initialization data, etc.).

## 2. Nagios, monitoring level and standard agent

### 2.1. Nagios

Nagios is an open source, unix-based software widely used to monitor networks and computers performance. It has a web-based front-end or console and provides monitoring of the network status (SMTP, POP3, SSH ãĀ̧e), host resources (CPU load, disk usage, memory load..) and services. Nagios controls hosts, servers and so on with native or custom plugins, and displays system status on a web-based GUI. Architecture is based on a client/server model: every host in the network runs a Nagios service (plugin) that sends information to the server. Nagios can be divided in three parts: the Scheduler, a server part that checks plugins at regular intervals (according to the configuration provided) and performs some actions based on their results; the GUI that is the web interface of Nagios, generated by GCI (Gateway Common Interface) and based on PHP technology, that allows the user to view status configuration, alerts, state buttons and so on; the Plugin, native of Nagios, that defines a set of basic commands, developed by the Nagios community or some professional users, and can include custom commands. Each plugin checks a service or host status, and returns a result to Nagios according to an interface. In addition, Nagios has a very strong notification system that handles events of services or hosts

## 2.2. Monitoring level

Nagios provides technologies for network and resource monitoring. Network monitoring is defined as the ability to monitor computer network, health status, performance and so on. Resource monitoring is the ability to monitor and control allocated resources for each host. It is also important to evaluate computer performance. A combination of network and resources monitoring provides a generic-level monitoring, that is useful together the general trend of the network and general health status of the machines. More specific monitoring is possible thanks to the high flexibility of Nagios, that allows to develop custom checks to gather the status of a process. In this way a deeper and accurate view of the system, where specific process resources are involved, can be achieved. Finally, in order to reach an even wider system monitoring, it is possible to check the validity of operations (e.g. coordinate conversion): this level of monitoring requires a deeper knowledge of the single element, however. For SKA Telescope Manager a combination of generic-level monitoring and process status has been preferred.

## 2.3. TM.LMC Standard Agent

In order to achieve such a combination, a TM.LMC ([1], 2016) standard agent has been developed, extending NRPE (Nagios Remote Plugin Executor). It is a daemon, installed on the local machine, that gathers system metrics from applications and sends them back to the Nagios Server. It is composed by a collection of Nagios standard checks (such as CPU Load, RAM usage and so on) and custom scripts which read data from the TM applications. The overall data are aggregated and sent to Nagios Server, according to defined policies. The correctness of the application status sent to the Standard Agent is a responsibility of the TM Developer.

## 3. Lifecycle manager integration

In SKA TM, lifecycle management is the ability to control a software application in the following phases of its lifetime: configuration, start, stop/kill, update, upgrade or downgrade (version control). All these phases are managed by an IT automation software. The software configuration management system selected during TM.LMC design phase is Chef. It provides a set of tools to automate an infrastructure, basing upon the cookbook concept, the fundamental unit of configuration and policy distribution. A cookbook defines a scenario and contains everything required to support it. Each cookbook is composed by a set of recipes that specify the resources to use and the order they are applied, and contains attribute values, file distributions, templates and extensions to Chef, such as libraries, definitions and custom resources. The monitoring configuration is also part of the lifecycle management and must be automated because Nagios is deeply based on files. Chef can be used to do that and to create a workflow to deploy configuration files inside the network machines and the Nagios server. In this way it will be possible to create a dynamic monitoring configuration during every lifecycle step. TM Developers decide what needs to be monitored and this must happen every time a new release comes up. This information is loaded into the lifecycle repository so that the NPRE daemon can start its monitoring job with all the configuration files synchronized with the repository before the application starts.

## 4. TANGO Controls integration

The TANGO control system is a free open source device-oriented controls toolkit based on CORBA for controlling any kind of hardware or software. It also provides tools to build control and integrated systems. TANGO has been adopted at SKA level as a standard bus communication for all SKA Elements (e.g. Dishes, Infrastructure, TM ...). To support the large number of Elements across the SKA (each Element being provided with its own LMC), the TM wants to ensure hierarchical reporting in the Element LMC, which means that it has to provide the rolled-up health status, state and mode reporting for that Element. TM.LMC, as any other LMC, needs to report the monitoring information to Telescope Management (TM.TELMGT). In order to do that Nagios needs to be integrated in the TANGO facility. Nagios allows several ways to implement the communication with TANGO. An example is to read internal files that Nagios keep updated (for instance status.dat), but this way is unwieldy. It is also possible to install some useful plugins like NDOUtils which consists of a standalone daemon, a Nagios event broker, a several helper utilities and allows to export current and historical data from one or more Nagios instances to a MySQL database. The data stored in the database can be read from a Tango Device Server and exposed to TELMGT through dynamic at-tribute. Another way is to create a query using Nagios Json CGI, that provides an API to query object, status, and historical information through GET requests. Once queried ,they return valid JSON that can be parsed into JavaScript objects for client side models and processing. A Tango Device Server has been developed with Nagios Json CGIand using a particular feature of TANGO that is the dynamic attributes configured with an xml file. The separation among generic-level plus process status and correctness of operations essentially, defines the monitoring boundaries between TM.LMC and the other TM sub-elements. In fact, a TM application handles internally its exception/error and determines an aggregate status to communicate to the standard agent. The standard agent does not know the internal structure of the application but just its aggregate status, so that managing his error/exception is a responsibility of the TM application.

## 5. Fault Management

The Fault Management uses Nagios in order to perform its duty, as follows: 1. Detection, that is the ability to understand if a fault is occurring in the system; 2. Isolation, that is the ability to isolate a fault understanding where it is occurring; 3. Recovery, that is the ability to recover the situation ([2], 2016). A monitoring activity of Nagios together with alarm filtering creates the detection activity. The same activity together with log information creates the isolation while the recovery is essentially a control operation that TM.LMC can do: an online action, which is a lifecycle command (reconfigure, restart, etc.) or an offline activity, like raising a modification request for the software maintenance. Since the lifecycle management is made up with the help of an IT automation tool, it allows not only to perform a lifecycle operation on the application, but also to execute any kind of script directly into the server machine (where the TM application runs). Therefore, a TM developer could create a specific recovery procedure (that is a specific script stored in the Lifecycle Manager Repository) that can be executed by Nagios in case of a specific alarm condition.

## 6. Logging

Logging is an important component of the development cycle. It offers several advantages and provides precise context about the application run. Once in the code, the generation of logging output requires no human intervention and can be saved in a persistent medium to be studied at a later stage. The TM logging architecture is based on the ELK stack (Elasticsearch, Logstash and Kibana). Each element of the telescope sends the logs to the server that has the responsibility to store the data depending on their priority. Nagios, via its libraries, allows to forward log message using technologies chosen with standard structured messages. Nagios allows also to develop a check that retrieve log data information from the server and obtains metric performance.

## 7. Conclusion

Nagios is one of the most widely used systems in IT infrastructures monitoring. Its features make it preferable (with respect to other software packages) especially for the SKA Project. SKA.TM, in particular, will be a large and complex environment composed by several software applications, integrated by TANGO, working in orchestrated way: Nagios, thanks to its flexibility, appears especially suitable for integrating the monitoring software in this heterogeneous environment.